\documentclass[journal]{IEEEtran}

\usepackage{cite}          
\usepackage{amsmath,amssymb,amsfonts}  
\usepackage{graphicx}      
\usepackage{textcomp}      
\usepackage{algorithm}     
\usepackage{algorithmic}   
\usepackage{booktabs}      
\usepackage{multirow}      
\usepackage{color}         
\usepackage{url}           
\usepackage{tabularx}
\usepackage{comment} 
\usepackage[colorlinks=true, citecolor=blue, linkcolor=blue, urlcolor=blue]{hyperref}

\makeatletter
\AtBeginEnvironment{algorithmic}{%
    \let\refstepcounter\H@refstepcounter
}
\makeatother

\usepackage[caption=false,font=footnotesize]{subfig}

\usepackage[table]{xcolor}
\usepackage{multirow,booktabs}
\usepackage{array}
\definecolor{hdryellow}{HTML}{FEF9E7}   
\definecolor{oursgreen}{HTML}{EAF5EB}   

\DeclareMathOperator*{\argmax}{argmax}

\begin{document}

\title{A-MADiff: Attention-Guided Multi-Agent DRL with Diffusion Policies for Memory-Aware Task Orchestration in Mobile AIGC Networks}

\author{Chongzhi~Wu,
        Zhengtao~Li,
        Jiawen~Kang,~\IEEEmembership{Senior Member,~IEEE,}
        Jinbo~Wen,
        Xiaohuan~Li,~\IEEEmembership{Member,~IEEE,}
        Maomao~Zhang,
        and~Ekram~Hossain,~\IEEEmembership{Fellow,~IEEE}
\thanks{C. Wu, Z. Li, and J. Kang are with the School of Automation, Guangdong University of Technology, Guangzhou 510006, China (e-mails: wu2261952901@mails.gdut.edu.cn; lizhengtao@mails.gdut.edu.cn; kavinkang@gdut.edu.cn). J. Wen is with the Department of Computer Science, City University of Hong Kong, Hong Kong (e-mail: jinbo1608@163.com). X. Li is with the School of Information and Communication, Guilin University of Electronic Technology, Guilin 541004, China (e-mail: lxhguet@guet.edu.cn). M. Zhang is with the Anhui Engineering Research Center for Agricultural Product Quality Safety Digital Intelligence, Fuyang Normal University, Fuyang 236037, China, and also with the School of Physics and Electronic Engineering, Fuyang Normal University, Fuyang 236037, China (e-mail: zhangmaohi@163.com). E. Hossain is with the Department of Electrical and Computer Engineering, University of Manitoba, Winnipeg, MB R3T 5V6, Canada (e-mail: Ekram.Hossain@umanitoba.ca).}
\thanks{\textit{Corresponding author: Jiawen Kang.}}}


\maketitle

\begin{abstract}
Artificial Intelligence-Generated Content (AIGC) services employ Generative AI (GenAI) models to automatically generate diverse content. Mobile AIGC networks host GenAI models on edge-located AIGC Service Providers (ASPs) to deliver low-latency and personalized AIGC services for mobile users. However, AIGC inference tasks typically occupy GPU memory until task completion, causing GPU memory exhaustion at serving ASPs and triggering out-of-memory failures rather than merely increasing service latency. Existing studies on AIGC task orchestration have largely overlooked GPU memory feasibility constraints. To address this issue, we develop a cooperative multi-agent orchestration framework, in which each edge node is equipped with a scheduling agent to route tasks to local ASPs or neighboring edge nodes. Since scheduling agents make decisions based only on local observations, while peer offloading couples their resource states and long-term utilities, we formulate the orchestration process as a cooperative Decentralized Partially Observable Markov Decision Process (Dec-POMDP). To solve the Dec-POMDP, we propose an \underline{A}ttention-guided \underline{M}ulti-\underline{A}gent deep reinforcement learning algorithm with \underline{Diff}usion policies (A-MADiff) under the centralized training with a decentralized execution paradigm. A-MADiff employs diffusion-based decentralized actors to generate multi-modal preferences over feasible orchestration actions, and an attention-guided centralized critic to estimate per-agent values from cross-agent states under GPU memory heterogeneity. Numerical results demonstrate that A-MADiff significantly improves the cumulative reward 
over the state-of-the-art baseline.
\end{abstract}


\IEEEpeerreviewmaketitle


\section{Introduction}\label{sec:introduction}

Artificial Intelligence-Generated Content (AIGC) employs Generative AI (GenAI) models to automatically produce digital content such as text, images, audio, and video, and has expanded from single-modality generation to multi-modal collaborative synthesis~\cite{cao2025survey, xu2024unleashing}. Advances in GenAI models, such as Large Language Models (LLMs) and diffusion models, have substantially improved the content quality and task adaptability of AIGC services~\cite{brown2020language, ho2020denoising, rombach2022high}, accelerating the transition toward real-time, request-driven generation for interactive mobile applications. However, since GenAI model inference on resource-constrained mobile devices incurs prohibitive computational load, memory demand, and processing latency~\cite{xu2023sparks, qu2025mobile, lai2024resource}, Mobile Edge Computing (MEC) provides a feasible paradigm for GenAI model inference. By deploying GenAI models to AIGC Service Providers (ASPs) at the network edge, mobile users can access low-latency and personalized AIGC services~\cite{xu2024unleashing}, such as text-to-image generation for smartphone users~\cite{du2024diffusion}, avatar synthesis for the mobile metaverse~\cite{xu2023sparks}, and LLM-based conversational assistants~\cite{qu2025mobile}. Recent studies have investigated mobile AIGC service optimization from several perspectives, including ASP selection~\cite{du2024diffusion}, multi-service orchestration~\cite{liu2025qos}, device-edge collaborative offloading~\cite{zhou2024deploying, xu2026joint}, resource coordination and management~\cite{liu2024joint, lai2024resource}, and incentive mechanism design~\cite{wen2024diffusion}. These works show that achieving high-quality mobile AIGC services requires jointly considering task requirements, model capabilities, edge resources, and communication conditions.

However, most existing studies on mobile AIGC service optimization have focused on computational resources, communication bandwidth, caching strategies, or service quality requirements~\cite{xu2024unleashing, lai2024resource, liu2024joint}, while the role of GPU memory constraints in task orchestration remains largely unexplored. In practice, GenAI model inference executes as massively parallel tensor computation on GPUs, and the model weights, Key-Value (KV) cache, and intermediate activations should stay resident in GPU memory throughout execution, imposing substantial memory overhead that neither Central Processing Unit (CPU) memory nor disk storage can accommodate~\cite{kwon2023efficient, li2024llm}. This overhead has been identified as a primary factor limiting system throughput and scalability in GenAI model inference~\cite{kwon2023efficient, li2024llm, prabhu2025vattention}. A fundamental asymmetry exists between GPU memory and computational resources in this context. Specifically, when computational capacity is insufficient, an AIGC task incurs longer execution latency but can still eventually be completed. However, when the GPU memory available at runtime falls short of the task's memory requirement, the task triggers an Out-of-Memory (OOM) failure and cannot be executed at all~\cite{xu2023sparks}. This distinction means that the matching between the memory requirements of AIGC tasks and the available GPU memory of serving ASPs constitutes a hard feasibility condition at the execution layer, rather than an ordinary soft resource cost that merely degrades service quality.

Moreover, the GPU memory constraint further amplifies the complexity of task orchestration in distributed mobile AIGC networks. AIGC requests may arrive dynamically at different edge nodes, and the ASPs managed by each edge node can differ in the inference rate, service capability, and GPU memory capacity. Under such conditions, each edge node relies on a software scheduler, referred to as a scheduling agent, to route arriving requests under partial observability~\cite{zhao2022multi}. AIGC task orchestration therefore becomes a cooperative orchestration problem among multiple scheduling agents rather than a single-entity ASP selection problem. Since GPU memory requirements, edge resource heterogeneity, queue dynamics, and cross-node task flow are tightly coupled~\cite{xu2024unleashing, qu2025mobile, zhao2022multi}, existing methods that center on a single scheduling entity for ASP selection or service orchestration do not account for these distributed interactions~\cite{du2024diffusion, liu2025qos}. The long-horizon cooperative decision process required by memory-constrained mobile AIGC networks has not been adequately studied.

This paper investigates GPU memory-aware task orchestration in resource-constrained and partially observable mobile AIGC networks. We first formulate the orchestration of local execution, peer offloading, and queue deferral under GPU memory feasibility constraints as a cooperative Decentralized Partially Observable Markov Decision Process (Dec-POMDP). To solve this problem, we propose A-MADiff, an \textbf{\underline{A}}ttention-guided \textbf{\underline{M}}ulti-\textbf{\underline{A}}gent Deep Reinforcement Learning (MADRL) algorithm with \textbf{\underline{Diff}}usion policies, following the Centralized Training with Decentralized Execution (CTDE) paradigm. Since the optimal action preference over these orchestration modes varies with the system state and can be multi-modal, A-MADiff employs diffusion-based decentralized actors that generate state-conditioned discrete action distributions through iterative reverse denoising. For cross-agent value estimation, A-MADiff further adopts an attention-guided centralized critic that adaptively weights multi-agent state information during centralized training, accounting for the non-uniform influence of neighboring nodes under GPU memory heterogeneity. The main contributions of this paper are summarized as follows:

\begin{itemize}
\item We propose a cooperative multi-agent task orchestration framework for mobile AIGC networks. In this framework, distributed scheduling agents collaboratively orchestrate dynamic AIGC requests under partial observability, while heterogeneous ASPs provide different inference capabilities and GPU memory capacities. In addition, peer offloading couples multiple edge nodes by interconnecting their task loads and resource states.

\item We develop a GPU memory-aware AIGC task processing model and formulate the dynamic task orchestration problem as a cooperative Dec-POMDP. The formulation explicitly incorporates the compatibility between GPU memory requirements and the available GPU memory resources at ASPs as a hard feasibility constraint for AIGC task execution. It further captures the temporal coupling among dynamic task arrivals, queue evolution, peer offloading, and long-term network utility.

\item To address the formulated cooperative Dec-POMDP, we propose A-MADiff, a novel attention-guided MADRL algorithm with diffusion policies. Following the CTDE paradigm, A-MADiff leverages diffusion-based decentralized actors to generate discrete orchestration action preferences from local observations. An attention-guided centralized critic then estimates agent-specific marginal action values by adaptively weighting cross-agent states during training. Numerical results demonstrate that A-MADiff outperforms the state-of-the-art baseline.
\end{itemize}



\section{Related Work}
\label{sec:related_work}

\subsection{Task Orchestration in Mobile AIGC Networks}
\label{subsec:rw_task_orchestration}

Research on mobile AIGC networks has evolved from model deployment toward request-driven service provisioning and task orchestration at the network edge~\cite{xu2024unleashing}. Subsequent studies have investigated AIGC request processing from various perspectives, including GenAI model caching and inference, model partitioning, AIGC task offloading, resource optimization, and incentive mechanism design~\cite{xu2023sparks, lai2024resource, zhou2024deploying, liu2024joint, liu2025optimizing, wen2024diffusion}. For instance, Du et al.~\cite{du2024diffusion} formulated dynamic ASP selection as a single-agent decision problem and proposed a diffusion-based DRL algorithm to generate selection strategies conditioned on ASP resource states and user-perceived utility. Liu et al.~\cite{liu2025qos} investigated AIGC service orchestration with heterogeneous task types. They further proposed an attention-diffusion-based DRL algorithm to derive effective orchestration strategies, where self-attention was integrated into the diffusion noise prediction network to capture correlations between task features and edge server states. Although existing studies provide foundations for service selection and request processing in mobile AIGC networks, they center on a single scheduling entity with full state observability. By contrast, this paper addresses multi-agent cooperative task orchestration under partial observability, where each scheduling agent decides among local execution, peer offloading, and queue deferral.

\subsection{GPU Memory Constraints in GenAI Model Inference}
\label{subsec:rw_gpu_memory}

Deploying GenAI models at the network edge typically incurs substantial GPU memory overhead due to model weights, KV cache, and intermediate activations~\cite{qu2025mobile}. In particular, the iterative denoising process of diffusion-based generators introduces additional sustained memory overhead. Studies on large-scale inference serving have identified GPU memory as a critical bottleneck limiting system throughput and scalability~\cite{kwon2023efficient,li2024llm,prabhu2025vattention}. Proposed solutions address this bottleneck through paged KV cache management~\cite{kwon2023efficient}, virtual-physical memory decoupling~\cite{prabhu2025vattention}, and optimized batching and scheduling~\cite{li2024llm}. On the edge inference side, model partitioning and multi-tier recursive offloading have been adopted to improve inference efficiency under heterogeneous resource constraints~\cite{zhang2025edgeshard, wu2026recursive, zeng2026h2o}. However, existing solutions primarily focus on the model serving or inference execution layer, optimizing AIGC task processing after task assignment to specific inference systems. In contrast, this work investigates memory-constrained task orchestration at the scheduling layer, where GPU memory feasibility is considered before task execution. Since insufficient GPU memory at runtime causes outright execution failure rather than mere latency increase~\cite{xu2023sparks}, we model the compatibility between task memory requirements and ASP memory availability as a hard feasibility constraint at the orchestration layer.

\subsection{MADRL with Diffusion Policies}
\label{subsec:rw_madrl_diffusion}

The cooperative multi-agent nature of the problem studied in this paper motivates the adoption of the CTDE paradigm introduced by Lowe et al.~\cite{lowe2017multi}, which serves as the foundation for cooperative MADRL methods~\cite{yu2022surprising} and their applications to edge computing task offloading~\cite{zhao2022multi, chen2026decentralized}. However, existing MADRL methods still face limitations in both components of the actor-critic architecture, i.e., single-pass actors lack the capability to characterize state-dependent multi-modal action preferences, while centralized critics typically fail to capture heterogeneous contributions from different agents. Iqbal et al.~\cite{iqbal2019actor} relieved the critic-side limitation with an attention mechanism that adaptively weights cross-agent information. This principle was later adopted for AIGC task offloading at the edge~\cite{li2024multi} and for cooperative UAV positioning~\cite{xu2026transformer}. Wang et al.~\cite{wang2023diffusion} addressed the actor-side limitation by modeling the policy as a conditional diffusion model, and demonstrated that iterative reverse denoising captures multi-modal action distributions more accurately than Gaussian or variational autoencoder policies. Diffusion-based DRL has been applied to network optimization tasks, including ASP selection, AIGC service orchestration and scheduling, contract design, and low-carbon optimization~\cite{du2024enhancing, du2024diffusion, liu2025qos, xu2026eat, wen2026diffusion, wen2026hybridrag}, in which a single agent observes the full environment state, and attention serves only as a policy-side feature encoder~\cite{liu2025qos, xu2026eat}.

By contrast, multi-agent diffusion-based DRL remains at an early stage. Existing studies mainly learn generative policies or planners from offline datasets~\cite{zhu2024madiff, li2025dof}, estimate value distributions on multi-agent control benchmarks~\cite{zhong2026mad3pg}, or extract request features for routing decisions in cloud-edge LLM serving~\cite{yao2025enhancing}, none of which models GPU memory availability as a hard feasibility constraint on discrete orchestration actions in an online cooperative Dec-POMDP. Therefore, we extend diffusion policies to the online cooperative multi-agent setting under the CTDE paradigm and propose A-MADiff. Within A-MADiff, diffusion-based decentralized actors generate feasibility-masked discrete action preferences, and an attention-guided centralized critic estimates agent-specific action values.

\begin{figure}[t]
\centering
\includegraphics[width=0.45\textwidth]{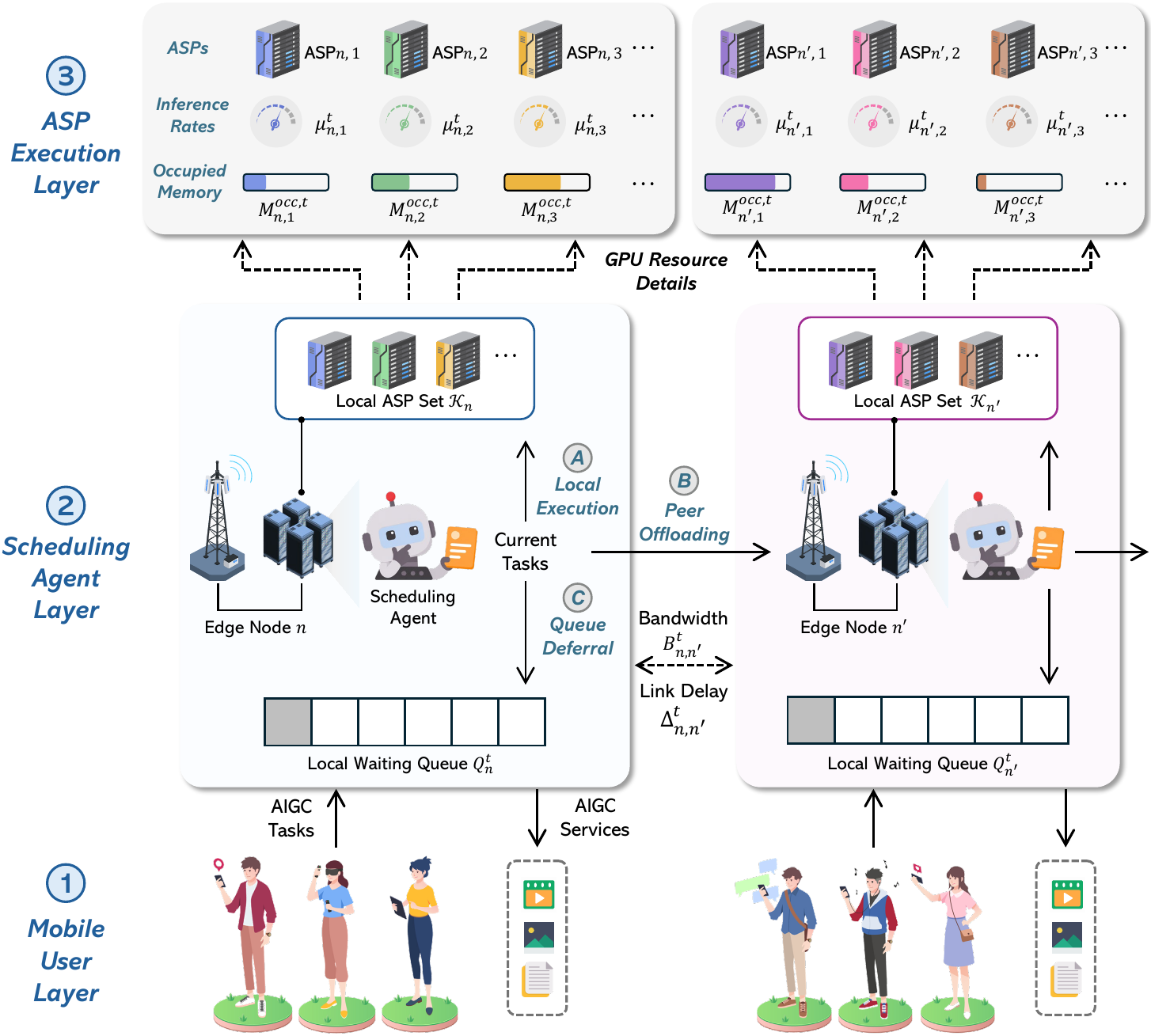}
\caption{Overview of the proposed GPU memory-aware task orchestration framework in distributed mobile AIGC networks.}
\label{fig:system_model}
\end{figure}


\section{GPU Memory-Aware Task Orchestration Framework}\label{sec:system_model}

\subsection{Framework for Mobile AIGC Task Orchestration}\label{subsec:architecture}

As illustrated in Fig.~\ref{fig:system_model}, we consider a distributed mobile AIGC network in which mobile users request AIGC services. Since mobile devices cannot independently sustain the computational and memory overhead of GenAI model inference, AIGC requests are uploaded to nearby edge nodes and executed by ASPs deployed on edge servers. Each edge node is an edge computing site co-located with a base station or access point, and it accommodates multiple edge servers, where each edge server hosts one ASP that runs a pre-trained GenAI model on a dedicated GPU. Mobile users move across the coverage areas of edge nodes, and thus the serving edge node of a user changes over time. Since this change is much slower than the execution of a single AIGC inference task, mobility manifests as time-varying request arrivals at each edge node. A software scheduler (i.e., a scheduling agent) deployed at each edge node manages the local task flow. The set of scheduling agents is denoted by $\mathcal{N} = \{1, \ldots, N\}$, where $N$ is the total number of scheduling agents. Each scheduling agent $n \in \mathcal{N}$ manages a set of local ASPs, denoted by $\mathcal{K}_n$, and routes each incoming task to a local ASP, a neighboring scheduling agent, or the local waiting queue. Since ASPs may host GenAI models of different types or scales on different GPU hardware, they can differ in the inference rate, service capability, and GPU memory capacity. The specific ASP admission and placement are handled by a unified execution-layer rule based on the GPU memory available at runtime. 

Scheduling agents can transfer tasks among each other through edge-side connections~\cite{zhao2022multi, wu2026recursive, chen2026decentralized}. The set of neighboring scheduling agents of agent $n$ is denoted by $\mathcal{N}_n$. For an AIGC task arriving at scheduling agent $n$, agent $n$ can assign the task to a local ASP for execution, migrate it to a neighboring scheduling agent $n' \in \mathcal{N}_n$, or defer it to the local waiting queue. If a task is migrated to a neighboring scheduling agent, it first enters the waiting queue at the target node and participates in subsequent scheduling decisions at later decision steps. This paper focuses on GPU memory feasibility modeling at the task orchestration layer, rather than on low-level GPU memory allocator design, memory fragmentation management, KV cache offloading, or model partitioning mechanisms.

\subsection{Dynamic Request Arrival, Queue, and Decision Steps}\label{subsec:arrival_queue}

At each scheduling agent $n$, AIGC requests arrive dynamically following a stochastic process with arrival rate $\lambda_n$~\cite{du2024diffusion, liu2025qos, zhao2022multi}. The $i$-th AIGC task is defined as
\begin{equation}\label{eq:task_tuple}
\tau_i = (t_i^{\mathrm{arr}}, w_i, m_i, D_i, h_i),
\end{equation}
where $t_i^{\mathrm{arr}}$ denotes the task arrival time, $w_i$ denotes the AIGC inference workload, $m_i$ denotes the GPU memory required for execution, $D_i$ denotes the data payload to be transmitted during cross-node migration, and $h_i$ denotes the number of offloading hops the AIGC task has already undergone.

For scheduling agent $n$ at decision step $t$, we denote the current task awaiting orchestration as $c_n^t$. When a new AIGC request arrives, the new AIGC task becomes $c_n^t$ if the scheduling agent currently has no pending tasks. Otherwise, the request enters the local waiting queue. Each scheduling agent maintains a finite-capacity waiting queue $\mathbf{Q}_n^t$ with length $q_n^t = |\mathbf{Q}_n^t|$, subject to $q_n^t \leq Q_{\max}$, where $Q_{\max}$ denotes the maximum waiting queue capacity. The waiting queue stores deferred tasks, tasks migrated from neighboring scheduling agents, and locally arrived tasks that have not yet been processed. When the queue reaches its maximum capacity, no additional tasks can enter.

To clarify the relationship between execution latency and decision steps, this paper adopts a time-slotted decision model with discrete scheduling and continuous execution time~\cite{zhao2022multi, chen2026decentralized}. The duration of each decision step is normalized to $\Delta t = 1$. At the beginning of step $t$, the system first releases GPU memory occupied by completed tasks based on the current global time. Each scheduling agent then makes a local execution, peer offloading, or queue deferral decision for the current task. After all agents complete their actions, the global time advances to $t + 1$, newly arrived tasks within the elapsed window enter the system, and the next tasks are dequeued for orchestration. When task $i$ begins execution at decision step $t$ on ASP $k \in \mathcal{K}_n$, the task completes at time $t + T_{n,k}^{\mathrm{exe}}(i, t)$, where $T_{n,k}^{\mathrm{exe}}(i, t)$ denotes the execution latency of task $i$ on ASP $k$. The task continuously occupies the corresponding GPU memory until completion.

\subsection{ASP Execution and GPU Memory Occupancy}\label{subsec:asp_execution}

For ASP $k \in \mathcal{K}_n$ managed by scheduling agent $n$, let $\mu_{n,k}^t$ denote its effective inference rate at decision step $t$ and $M_{n,k}^{\mathrm{cap}}$ denote its total GPU memory capacity. When task $i$ is assigned to ASP $k$, its execution latency is given by
\begin{equation}\label{eq:exec_latency}
T_{n,k}^{\mathrm{exe}}(i, t) = \frac{w_i}{\mu_{n,k}^t},
\end{equation}
which indicates that a higher effective inference rate yields a shorter execution latency for a given task. AIGC inference tasks continuously occupy GPU memory during execution. We denote the set of AIGC tasks currently running on ASP $k$ managed by scheduling agent $n$ at decision step $t$ as $\mathcal{C}_{n,k}^t$. Since each running task $i' \in \mathcal{C}_{n,k}^t$ occupies the GPU memory $m_{i^\prime}$, the occupied and available GPU memory of ASP $k$ are respectively expressed as
\begin{equation}\label{eq:gpu_memory}
M_{n,k}^{\mathrm{occ},t} = \!\!\!\sum_{i' \in \mathcal{C}_{n,k}^t}\!\!\! m_{i'}, \quad
M_{n,k}^{\mathrm{av},t} = M_{n,k}^{\mathrm{cap}} - M_{n,k}^{\mathrm{occ},t}.
\end{equation}
When task $i^{\prime}$ is completed, its occupied GPU memory is released, and the task is removed from $\mathcal{C}_{n,k}^t$. Consequently, the available GPU memory of ASP $k$ changes dynamically as tasks begin execution and complete. Hence, the GPU memory feasibility condition for task $i$ to begin execution on ASP $k$ at step $t$ is $M_{n,k}^{\mathrm{av},t} \geq m_i$. Unlike the effective inference rate, which primarily affects execution latency, GPU memory availability directly determines whether a task can execute on the target ASP at runtime.

\subsection{AIGC Task Processing and Cross-Node Flow}\label{subsec:task_processing}

For the current task $c_n^t$ at scheduling agent $n$ during decision step $t$, the system determines the processing mode based on local ASP states, neighboring node conditions, and waiting queue capacity. A task can be accepted and executed by a local ASP, migrated to the waiting queue of a neighboring scheduling agent, or deferred to the local waiting queue for subsequent scheduling. Different processing modes cause changes to the running task sets, GPU memory occupancy, and waiting queue states.

\subsubsection{Local execution}\label{subsubsec:local_exec}
When task $i$ is processed locally at scheduling agent $n$, the system selects an eligible ASP from the local ASP set $\mathcal{K}_n$. The set of runtime-admissible local ASPs for task $i$ is defined as
\begin{equation}\label{eq:runtime_admissible}
\mathcal{K}_n^{\mathrm{run}}(i, t) = \{k \in \mathcal{K}_n \mid M_{n,k}^{\mathrm{av},t} \geq m_i\}.
\end{equation}
When $\mathcal{K}_n^{\mathrm{run}}(i, t) \neq \emptyset$, task $i$ can be assigned to one of the runtime-admissible ASPs for execution. If this set is empty, no local ASP has sufficient available GPU memory at runtime to possess task $i$, and local execution fails the admission check.

\subsubsection{Peer offloading}\label{subsubsec:offloading}
When task $i$ is not executed locally, scheduling agent $n$ can migrate it to a neighboring scheduling agent $n' \in \mathcal{N}_n$. Task migration requires transmitting the task data payload over wired backhaul links between edge nodes, and the resulting communication latency is
\begin{equation}\label{eq:comm_latency}
T_{n,n'}^{\mathrm{comm}}(i, t) = \frac{D_i}{B_{n,n'}^t} + \Delta_{n,n'}^t,
\end{equation}
where $B_{n,n'}^t$ denotes the transmission bandwidth and $\Delta_{n,n'}^t$ denotes the link latency between scheduling agents $n$ and $n'$. After migration, the task enters the waiting queue $\mathbf{Q}_{n'}^t$ of the target scheduling agent and is scheduled by that agent in subsequent decision steps.

Peer offloading is constrained by the hop count and the target queue capacity. The offloading hop count of task $i$ needs to satisfy $h_i < H_{\max}$, where $H_{\max}$ denotes the maximum number of allowed offloading hops. Otherwise, the task is no longer eligible for further migration. The target scheduling agent $n'$ also needs to have remaining queue capacity, i.e., $q_{n'}^t < Q_{\max}$. If the target queue is full, the task cannot be accepted by the target node. If the task successfully enters the target queue, its hop count is incremented, and it becomes part of the target node's pending task flow.

\subsubsection{Queue deferral}\label{subsubsec:deferral}
If task $i$ is neither accepted by a local ASP nor migrated to a neighboring scheduling agent, scheduling agent $n$ can defer it to the local waiting queue $\mathbf{Q}_n^t$ so that the task participates in scheduling at subsequent decision steps~\cite{han2023impatient}. However, if the local queue length has already reached the maximum capacity $Q_{\max}$, the current task cannot enter the local waiting queue.

The above processing modes cause the system state to evolve continuously. Local execution changes the target ASP's running task set and GPU memory occupancy. Peer offloading changes the target scheduling agent's waiting queue state, and queue deferral changes the local waiting queue state. As a result, the outcome of current task processing affects the available GPU memory, queue load, and task admission conditions at subsequent decision steps.


\section{Problem Formulation}\label{sec:problem_formulation}

Mobile AIGC task orchestration involves online decisions made by multiple scheduling agents under partial information. Each scheduling agent can access only local task, queue, ASP resource, and neighboring node state summaries, rather than the complete operating states of all agents across the network. Meanwhile, the outcome of current task processing propagates to subsequent decision steps through changes in queue load, available GPU memory, and cross-node task flow. Since the decisions of multiple scheduling agents jointly determine the AIGC task processing performance, we formulate the memory-constrained mobile AIGC task orchestration problem as a cooperative Dec-POMDP, represented by the tuple
\begin{equation}\label{eq:dec_pomdp}
\mathcal{M} = \langle \mathcal{N}, \mathcal{S}, \{\mathcal{O}_n\}_{n \in \mathcal{N}}, \{\mathcal{A}_n\}_{n \in \mathcal{N}}, \mathcal{P}, \Omega, R, \gamma \rangle,
\end{equation}
where $\mathcal{S}$ is the environment state space, $\mathcal{O}_n$ is the local observation space of scheduling agent $n$, $\mathcal{A}_n$ is the action space of scheduling agent $n$, $\mathcal{P}$ is the true state transition function, $\Omega$ is the observation function, $R$ is the cooperative reward function, and $\gamma \in (0, 1)$ is the discount factor. We distinguish the environment state from the centralized training input used by the critic. The state describes the Markov transition dynamics of the environment, while the critic uses the joint local observations as a partial-state representation for centralized function approximation during training.

\subsection{Environment State and Local Observation}\label{subsec:state_obs}

At decision step $t$, the environment state is defined as
\begin{equation}\label{eq:true_state}
S^t = \left( \mathbf{c}^t,\; \mathbf{Q}^t,\; \mathbf{X}^t,\; \mathbf{L}^t,\; \mathbf{R}^t \right),
\end{equation}
where $\mathbf{c}^t = \{c_n^t\}_{n \in \mathcal{N}}$ collects the current tasks at all~scheduling agents, and $\mathbf{Q}^t = \{\mathbf{Q}_n^t\}_{n \in \mathcal{N}}$ collects all waiting queues together with the attributes of stored tasks. The term $\mathbf{X}^t = \{\mu_{n,k}^t, M_{n,k}^{\mathrm{cap}}, M_{n,k}^{\mathrm{occ},t}, M_{n,k}^{\mathrm{av},t}, \mathcal{C}_{n,k}^t\}$ aggregates the ASP inference rates, GPU memory states, and running task sets, where $\mathcal{C}_{n,k}^t$ includes running tasks and their remaining completion times. The link state set $\mathbf{L}^t = \{B_{n,n'}^t, \Delta_{n,n'}^t\}$ consists of the bandwidth $B_{n,n'}^t$ and link latency $\Delta_{n,n'}^t$ between neighboring agents, and $\mathbf{R}^t$ represents the stochastic state governing task arrival realizations and resource perturbations. Given a joint action, $S^t$ contains all information required to determine queue dynamics, GPU memory release, and task flow, thereby satisfying the Markov state transition property.

During decentralized execution, scheduling agent $n$ makes decisions based on its local observation $o_n^t$. This observation consists of the current task feature $\phi_n^t$, the local state summary $\psi_n^t$, and the neighboring state summary $\chi_n^t$, given by
\begin{equation}\label{eq:local_obs}
o_n^t = \left( \phi_n^t,\; \psi_n^t,\; \chi_n^t \right).
\end{equation}
When the pending task at agent $n$ is the $i$-th AIGC task $\tau_i$, the task feature is $\phi_n^t = (w_i, m_i, D_i, h_i)$. If no task is pending, $\phi_n^t$ is a zero vector. We define the local state summary as
\begin{equation}\label{eq:local_summary}
\psi_n^t = \left( \rho_n^{\mathrm{comp},t},\; \rho_n^{\mathrm{mem},t},\; q_n^t / Q_{\max} \right),
\end{equation}
where the local computation load summary and the available GPU memory ratio summary are respectively given by
\begin{subequations}\label{eq:local_summaries}
\begin{align}
\rho_n^{\mathrm{comp},t} &= \frac{1}{|\mathcal{K}_n|} \sum_{k \in \mathcal{K}_n} \frac{|\mathcal{C}_{n,k}^t|}{1 + |\mathcal{C}_{n,k}^t|}, \label{eq:comp_load}\\
\rho_n^{\mathrm{mem},t} &= \frac{1}{|\mathcal{K}_n|} \sum_{k \in \mathcal{K}_n} \frac{M_{n,k}^{\mathrm{av},t}}{M_{n,k}^{\mathrm{cap}}}. \label{eq:mem_ratio}
\end{align}
\end{subequations}
Moreover, the neighboring state summary is expressed as
\begin{equation}\label{eq:neighbor_summary}
\chi_n^t = \left\{ \bar{M}_{n'}^{\mathrm{av},t},\; \bar{\mu}_{n'}^t,\; \hat{\rho}_{n'}^t,\; \hat{\Delta}_{n,n'}^t \right\}_{n' \in \mathcal{N}_n},
\end{equation}
where $\bar{M}_{n'}^{\mathrm{av},t}$ represents the available GPU memory summary at the target node, $\bar{\mu}_{n'}^t$ represents the inference capability summary, $\hat{\rho}_{n'}^t$ represents the noisy load summary, and $\hat{\Delta}_{n,n'}^t$ represents the noisy link latency summary. The local observation is written as $o_n^t = \Omega_n(S^t)$, i.e., a local summary obtained by applying the observation function to the state. These summary observations do not include complete task queues or per-ASP running task sets at other nodes.

During training, the centralized critic uses the joint local observation as the centralized training representation:
\begin{equation}\label{eq:joint_obs}
\tilde{s}^t = \left[ o_1^t, o_2^t, \ldots, o_N^t \right].
\end{equation}
It is worth noting that this joint observation $\tilde{s}^t$ serves as a partial-state representation for function approximation and does not represent the complete true state $S^t$.

\subsection{Action Space and Structurally Feasible Action Set}\label{subsec:action_space}

At each decision step, scheduling agent $n$ selects one orchestration action for the current task. The action space consists of three categories: local execution, peer offloading, and queue deferral, defined as
\begin{equation}\label{eq:action_space}
\mathcal{A}_n = \{a_{\mathrm{loc}}\} \cup \{a_{n'} \mid n' \in \mathcal{N}_n\} \cup \{a_{\mathrm{def}}\},
\end{equation}
where $a_{\mathrm{loc}}$ assigns the current task to the local execution layer for admission, $a_{n'}$ migrates the current task to the neighboring scheduling agent $n'$, and $a_{\mathrm{def}}$ defers the current task to the local waiting queue. Since the specific ASP selection is delegated to the ASP admission and placement process, the action space of agent $n$ does not additionally enumerate all local ASPs, thereby avoiding action space expansion that scales with the number of ASPs.

We define $\mathcal{F}_n^t \subseteq \mathcal{A}_n$ as the structurally feasible action set of scheduling agent $n$ at decision step $t$. When $c_n^t = \tau_i$, the set of local ASPs whose physical GPU memory capacity can accommodate task $i$ is first defined as
\begin{equation}\label{eq:kcap}
\mathcal{K}_n^{\mathrm{cap}}(i) = \left\{ k \in \mathcal{K}_n \;\middle|\; M_{n,k}^{\mathrm{cap}} \geq m_i \right\}.
\end{equation}
This set reflects only whether the GPU memory requirement of the task exceeds the physical GPU memory capacity of a local ASP, without using the available GPU memory at runtime as a filtering condition. Based on the above, the structural feasibility conditions for each action category are
\begin{equation}\label{eq:feasibility}
\begin{aligned}
a_{\mathrm{loc}} \in \mathcal{F}_n^t &\iff \mathcal{K}_n^{\mathrm{cap}}(i) \neq \emptyset, \\
a_{n'} \in \mathcal{F}_n^t &\iff n' \in \mathcal{N}_n,\; h_i < H_{\max},\; q_{n'}^t < Q_{\max}, \\
a_{\mathrm{def}} \in \mathcal{F}_n^t &\iff q_n^t < Q_{\max}.
\end{aligned}
\end{equation}
The local execution action only requires that at least one local ASP can physically host the task in terms of GPU memory capacity. The peer offloading action requires that the target node belongs to the neighboring set, the task has not exceeded the maximum hop count, and the target queue can still accept the task. The queue deferral action requires that the local waiting queue retains remaining capacity.

The structurally feasible action set is used to exclude actions that are definitively infeasible at the physical capacity, hop count, and queue capacity levels. The available GPU memory $M_{n,k}^{\mathrm{av},t}$ is not used to preemptively filter the local execution action. If scheduling agent $n$ selects $a_{\mathrm{loc}}$, the system performs the runtime admission check based on~(\ref{eq:runtime_admissible}). If $\mathcal{K}_n^{\mathrm{run}}(i, t) = \emptyset$, local execution fails due to insufficient GPU memory at runtime, and a failure penalty is returned to the policy. This design avoids giving the policy perfect knowledge of per-ASP memory availability at runtime, while ensuring that the GPU memory constraint takes effect as a hard feasibility condition.

\subsection{Immediate Reward Function}\label{subsec:reward}

The immediate reward captures the service benefit and scheduling cost of the current orchestration action. This paper follows the human-aware utility modeling approach used in AIGC service selection~\cite{du2024diffusion}. Specifically, when task $i$ is executed by ASP $k$ managed by scheduling agent $n$, the normalized service utility is defined as
\begin{equation}\label{eq:utility}
u_{n,k}^i = G\!\left( F_{n,k}(\tau_i) \right),
\end{equation}
where $F_{n,k}(\cdot)$ represents the generative service function of the GenAI model deployed on ASP $k$, and $G(\cdot)$ is the content quality or user-perceived utility evaluation function.

\subsubsection{Local execution reward}
When scheduling agent $n$ selects $a_{\mathrm{loc}}$, the system performs the admission check based on the runtime-admissible ASP set $\mathcal{K}_n^{\mathrm{run}}(i, t)$ defined in (\ref{eq:runtime_admissible}). If $\mathcal{K}_n^{\mathrm{run}}(i, t) \neq \emptyset$, the execution layer selects the ASP that maximizes the immediate local execution benefit:
\begin{equation}\label{eq:asp_select}
k_n^{\star}(i, t) = \argmax_{k \in \mathcal{K}_n^{\mathrm{run}}(i, t)} \left[ u_{n,k}^i - \eta_T \, T_{n,k}^{\mathrm{exe}}(i, t) \right],
\end{equation}
where $\eta_T > 0$ denotes the latency weight that converts execution and communication latency into an equivalent utility cost. This rule closes the ASP admission and placement process following the local execution action. The corresponding local execution reward is given by
\begin{equation}\label{eq:reward_loc}
r_{n,\mathrm{loc}}^t = u_{n,k_n^{\star}}^i - \eta_T \, T_{n,k_n^{\star}}^{\mathrm{exe}}(i, t).
\end{equation}
If $\mathcal{K}_n^{\mathrm{run}}(i, t) = \emptyset$, local execution fails due to insufficient GPU memory, and the agent receives the failure penalty $R_{\mathrm{fail}} < 0$.

\subsubsection{Peer offloading reward}
When a task is accepted by the target scheduling agent $n'$ through peer offloading, it enters the target's waiting queue rather than being immediately executed on a target ASP. Accordingly, the offloading reward is based on an estimated routing utility at the target node, with the cross-node communication cost and estimated subsequent processing cost subtracted, which is expressed as
\begin{equation}\label{eq:reward_off}
r_{n,\mathrm{off}}^t(n') = \hat{u}_{n'}^i - \eta_T \left[ T_{n,n'}^{\mathrm{comm}}(i, t) + \hat{T}_{n'}^{\mathrm{exe}}(i, t) \right],
\end{equation}
where $\hat{u}_{n'}^i = \hat{U}(\tau_i, \chi_{n'}^t)$ is the estimated routing utility computed from the target node's summary state, and $\hat{T}_{n'}^{\mathrm{exe}}(i, t) = \hat{T}(\tau_i, \chi_{n'}^t)$ is the estimated subsequent execution time based on the target node's current summary load.

It should be noted that (\ref{eq:reward_off}) captures the short-term shaping feedback for the current cross-node routing action, rather than the final completion reward of the task. After the task enters the target queue, its subsequent admission outcome and execution reward are captured by the state transitions and rewards at later decision steps. Therefore, the final completion reward of the same task is not double-counted during offloading.

\subsubsection{Queue deferral reward}
A deferred AIGC task does not generate immediate service utility. The scheduling agent instead incurs a fixed deferral cost $R_{\mathrm{def}} < 0$.

\subsubsection{Composite immediate reward}
Based on the selected action $a_n^t$, the immediate reward of scheduling agent $n$ at decision step $t$ is defined as
\begin{equation}\label{eq:reward_composite}
r_n^t =
\begin{cases}
r_{n,\mathrm{loc}}^t, & a_n^t = a_{\mathrm{loc}},\; \mathcal{K}_n^{\mathrm{run}}(i, t) \neq \emptyset, \\
r_{n,\mathrm{off}}^t(n'), & a_n^t = a_{n'},\; a_{n'} \in \mathcal{F}_n^t, \\
R_{\mathrm{def}}, & a_n^t = a_{\mathrm{def}},\; a_{\mathrm{def}} \in \mathcal{F}_n^t, \\
R_{\mathrm{fail}}, & \text{otherwise}.
\end{cases}
\end{equation}
For scheduling agents that have no task awaiting orchestration at the current step, the immediate reward is set to zero.

\begin{figure*}[!t]
\centering
\includegraphics[width=0.88\textwidth]{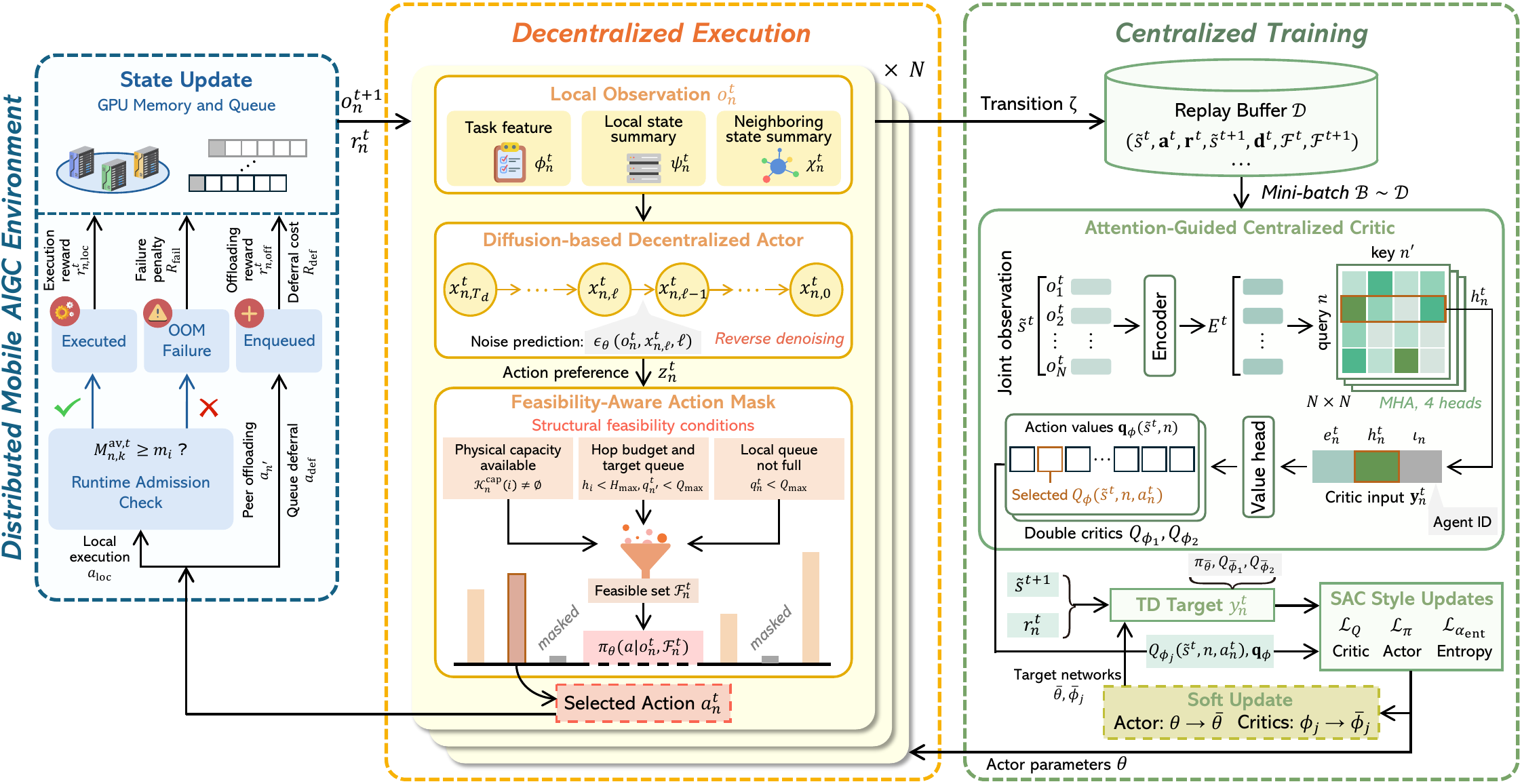}
\caption{Overall architecture of the proposed A-MADiff algorithm for GPU memory-aware AIGC task orchestration.}
\label{fig:framework}
\end{figure*}

\subsection{Cooperative Objective and CTDE Decomposition}\label{subsec:objective}

Task offloading and queue deferral create load coupling among different scheduling agents, so that a single agent's local reward does not fully reflect the impact of its decisions on the entire network. For instance, offloading a task to a neighboring node may alleviate local GPU memory pressure while increasing the target node's queue load. Queue deferral avoids immediate task dropping but can reduce the local reception capacity at subsequent decision steps. To capture the task processing performance of all scheduling agents, we adopt the cooperative reward, which is given by
\begin{equation}\label{eq:coop_reward}
r^t = \sum_{n \in \mathcal{N}} r_n^t.
\end{equation}

During the execution phase, each scheduling agent makes decentralized decisions based on its local observation $o_n^t$. The local policy of scheduling agent $n$ is denoted by $\pi_n(a_n^t \mid o_n^t)$, where $a_n^t$ is the selected orchestration action. The joint observation and joint action of all scheduling agents are denoted by $\mathbf{o}^t = \{o_n^t\}_{n \in \mathcal{N}}$ and $\mathbf{a}^t = \{a_n^t\}_{n \in \mathcal{N}}$, respectively. Hence, the joint decentralized policy composed of all local policies is
\begin{equation}\label{eq:joint_policy}
\pi(\mathbf{a}^t \mid \mathbf{o}^t) = \prod_{n \in \mathcal{N}} \pi_n(a_n^t \mid o_n^t).
\end{equation}

The objective is to learn a joint policy $\pi$ that maximizes the expected discounted cooperative return under dynamic task arrivals, queue dynamics, GPU memory occupancy changes, and cross-node task flow, which is expressed as
\begin{equation}\label{eq:objective}
\max_{\pi}\; J(\pi) = \mathbb{E}_{\pi, \mathcal{P}, \rho_0} \left[ \sum_{t=0}^{T-1} \gamma^t \, r^t \right],
\end{equation}
where $T$ denotes the number of decision steps per episode, and $\rho_0$ denotes the initial state distribution. Since $r^t = \sum_{n\in \mathcal{N}} r_n^t$, (\ref{eq:objective}) can equivalently be written as the sum of per-agent reward component returns. A-MADiff adopts the following normalized decomposition objective in its implementation:
\begin{equation}\label{eq:decomposed_obj}
J_{\mathrm{dec}}(\pi) = \frac{1}{N} \sum_{n=1}^{N} \mathbb{E}_{\pi, \mathcal{P}, \rho_0} \left[ \sum_{t=0}^{T-1} \gamma^t \, r_n^t \right].
\end{equation}
This decomposition does not alter the optimization direction but only rescales the objective magnitude. During training, the critic estimates the per-agent marginal action value, defined as the long-term return of agent $n$'s reward component when agent $n$ selects action $a$, given the centralized training representation and the current policies of all other agents:
\begin{equation}\label{eq:marginal_q}
Q_n^{\pi}(\tilde{s}, a) = \mathbb{E}_{\pi_{-n}, \mathcal{P}} \left[ \sum_{t' = t}^{T-1} \gamma^{t' - t} \, r_n^{t'} \;\middle|\; \tilde{s}^t = \tilde{s},\; a_n^t = a \right],
\end{equation}
where $\pi_{-n}$ denotes the policies of all agents except agent $n$, and $\tilde{s}^t$ is the centralized training representation composed of joint local observations. (\ref{eq:marginal_q}) shows that we adopt a per-agent marginal centralized critic rather than a joint-action critic $Q(S, \mathbf{a})$ that conditions on all agents' actions simultaneously. Since the joint action space grows as $|\mathcal{A}_n|^N$ and is computationally prohibitive at the scale of this problem, the critic instead integrates out other agents' actions under their current policies $\pi_{-n}$ and retains cross-agent coupling through the centralized training representation $\tilde{s}^t$. This design is consistent with the reward decomposition in~(\ref{eq:decomposed_obj}) and provides per-agent value estimates for cooperative policy optimization.


\section{A-MADiff for GPU Memory-Aware AIGC Task Orchestration}\label{sec:method}

As illustrated in Fig.~\ref{fig:framework}, A-MADiff addresses the multi-agent cooperative orchestration problem formulated in Section~\ref{sec:problem_formulation}. Each scheduling agent acts as a DRL agent that learns the orchestration policy from local observations. The policy of A-MADiff learns the task processing mode selection, while at the action execution time, the ASP admission and placement rule determines whether a specific ASP can accept the task based on its available GPU memory at runtime.

The memory awareness of A-MADiff is distributed across multiple stages rather than concentrated in a single mechanism. Before policy sampling, local GPU memory summaries enter the state representation, and a structural action mask removes physically infeasible actions. After action selection, the runtime admission check enforces GPU memory feasibility as a hard condition, and failure penalties feed back into the reward. Subsequent state transitions reflect the resulting GPU memory changes. This design avoids treating low-level GPU memory allocation as a direct agent output, while preserving the hard-constraint property of GPU memory at the execution layer.

\subsection{Diffusion Model-Based Decentralized Policy}\label{subsec:diffusion_actor}

In GPU memory-aware AIGC task orchestration, the quality of a scheduling action is jointly influenced by the task's GPU memory requirement, the ASP's available GPU memory, the queue state, the neighboring node load, and the communication latency. When local ASPs have sufficient available GPU memory resources, local execution may yield high immediate reward. When local resources are tight, or the queue is congested, peer offloading may be more beneficial. When current execution conditions are unstable, queue deferral can preserve the opportunity for subsequent scheduling. As a result, the optimal action preference distribution over the action space is state-dependent and not necessarily unimodal, requiring the policy to express complex conditional action distributions.

Standard single-pass policy networks with Gaussian or categorical outputs may not adequately capture such state-dependent multi-modal preferences. Diffusion-based policies address this limitation through iterative reverse denoising, which captures multi-modal distributions more accurately than single-pass alternatives and has been applied to network optimization and edge AIGC service selection~\cite{du2024enhancing, du2024diffusion, liu2025qos}. A-MADiff adopts this approach, employing a denoising diffusion model to generate discrete and state-conditioned action preferences from local observations, rather than replacing the ASP admission and placement rule.

Specifically, diffusion-based actors start from a Gaussian noise vector and perform multi-step reverse denoising conditioned on the local observation and the diffusion timestep, progressively refining the action logits. The multi-step denoising structure provides the policy network with a stronger state-conditioned action distribution representation than a standard feedforward actor. For scheduling agent $n$ with local observation $o_n^t$, A-MADiff first samples an initial noise vector $x_{n,T_d}^{t} \sim \mathcal{N}(\mathbf{0}, \mathbf{I})$, where $T_d$ is the number of reverse denoising steps, and $x_{n,T_d}^{t} \in \mathbb{R}^{|\mathcal{A}_n|}$ has the same dimension as the action space. We define $\beta_\ell$ as the noise schedule parameter at diffusion step $\ell$, $\alpha_\ell = 1 - \beta_\ell$, and $\bar{\alpha}_\ell = \prod_{j=1}^{\ell} \alpha_j$. During the reverse denoising process, the noise prediction network $\epsilon_\theta(\cdot)$ with parameters $\theta$ takes the local observation $o_n^t$, the current noisy vector $x_{n,\ell}^{t}$, and the diffusion step $\ell$ as input, and predicts the noise to be removed, expressed as
\begin{equation}\label{eq:noise_pred}
\hat{\epsilon}_{n,\ell}^{t} = \epsilon_\theta(o_n^t, x_{n,\ell}^{t}, \ell).
\end{equation}
Based on the predicted noise, the reverse transition mean is expressed as
\begin{equation}\label{eq:reverse_mean}
\mu_\theta(x_{n,\ell}^{t}, \ell, o_n^t) = \frac{1}{\sqrt{\alpha_\ell}} \left( x_{n,\ell}^{t} - \frac{\beta_\ell}{\sqrt{1 - \bar{\alpha}_\ell}} \hat{\epsilon}_{n,\ell}^{t} \right).
\end{equation}
A-MADiff then performs reverse sampling according to
\begin{equation}\label{eq:reverse_sample}
x_{n,\ell-1}^{t} = \mu_\theta(x_{n,\ell}^{t}, \ell, o_n^t) + \mathbb{I}_{\ell > 1} \sqrt{\tilde{\beta}_\ell} \, \xi,
\end{equation}
where $\xi \sim \mathcal{N}(\mathbf{0}, \mathbf{I})$ is sampled independently for each scheduling agent and denoising step, $\mathbb{I}_{\ell > 1}$ is the indicator function, and $\tilde{\beta}_\ell$ is the reverse process variance. After iterating from $T_d$ down to $1$, A-MADiff takes the final output $x_{n,0}^{t}$ as the action preference vector $z_n^t$ of scheduling agent $n$.

To ensure that the policy samples only from structurally feasible actions, A-MADiff normalizes the diffusion-generated action preference $z_n^t$ onto the structurally feasible action set:
\begin{equation}\label{eq:masked_policy}
\pi_\theta(a \mid o_n^t, \mathcal{F}_n^t) =
\begin{cases}
\dfrac{\exp(z_n^t(a))}{\sum_{\tilde{a} \in \mathcal{F}_n^t} \exp(z_n^t(\tilde{a}))}, & a \in \mathcal{F}_n^t, \\[6pt]
0, & a \notin \mathcal{F}_n^t.
\end{cases}
\end{equation}
When $\mathcal{F}_n^t = \emptyset$, A-MADiff uses a fallback mask $\{a_{\mathrm{def}}\}$ to form a computable distribution, and the environment determines whether this action can be accepted by the queue. If the queue cannot accept the task, a failure penalty is returned. (\ref{eq:masked_policy}) ensures that the policy sampling satisfies the structural constraints on physical capacity, hop count, and queue capacity. GPU memory insufficiency at runtime is not preemptively filtered by this mask, but rather determined by the runtime admission check after action execution.

The diffusion-based actors in A-MADiff are updated end-to-end through Soft Actor-Critic (SAC)-style policy improvement~\cite{haarnoja2018soft}, rather than using a standalone supervised denoising reconstruction objective. Specifically, the reverse denoising process serves as the action preference generator, and its parameters are optimized by the entropy-regularized actor loss through policy gradient signals. This is consistent with existing diffusion-based DRL methods~\cite{du2024diffusion,liu2025qos}, which use the diffusion model as a policy network guided by return signals from environment interaction rather than supervised labels.

\subsection{Attention-Guided Centralized Critic}\label{subsec:attention_critic}

In multi-agent task orchestration, the action of one scheduling agent affects the queue states, resource occupancy, and subsequent service capacity of other nodes. For example, offloading a task to a neighboring node increases the target node's waiting queue load and may reduce its capacity to accept future tasks. Consequently, the value estimation during training should incorporate cross-agent state information. However, not all scheduling agents contribute equally to the current decision. Typically, neighboring nodes with larger available GPU memory capacity, lower queue loads, or smaller communication latency have a stronger impact on the offloading value estimation of the task under consideration.

Standard centralized critics that concatenate all agents' observations treat each agent's contribution equally and cannot capture such non-uniform dependencies. Attention-based critics address this limitation by adaptively weighting cross-agent state information~\cite{iqbal2019actor}. Following this principle, A-MADiff constructs an attention-guided centralized critic that estimates the per-agent marginal action value defined in~(\ref{eq:marginal_q}). Unlike a joint-action critic $Q(S, \mathbf{a})$, this critic evaluates the long-term value of a single agent selecting a given local action under the centralized training representation and the current policies of other agents. This design is consistent with the reward decomposition objective in~(\ref{eq:decomposed_obj}) and improves scalability by avoiding the joint action space enumeration of size $|\mathcal{A}_n|^N$.

During training, the critic takes the joint local observation of all scheduling agents as the centralized training representation, as defined in~(\ref{eq:joint_obs}). The critic first applies a shared observation encoder $f_{\mathrm{enc}}(\cdot)$ to extract features from each scheduling agent's local observation, expressed as
\begin{equation}\label{eq:obs_encoding}
e_n^t = f_{\mathrm{enc}}(o_n^t), \quad n \in \mathcal{N}.
\end{equation}
The encoded results of all agents are stacked as $E^t = [e_1^t, e_2^t, \ldots, e_N^t]$. A-MADiff then applies Multi-Head Self-Attention (MHA) to model cross-agent state interactions:
\begin{equation}\label{eq:multi_head_attention}
H^t = \mathrm{MHA}(E^t, E^t, E^t),
\end{equation}
where the encoded matrix $E^t$ serves as the query, key, and value, and $H^t = [h_1^t, \ldots, h_N^t]$ denotes the attention context.

For each scheduling agent $n$, the critic takes the corresponding encoding $e_n^t$ and attention context $h_n^t$, and concatenates them with the one-hot agent identity vector $\iota_n$, yielding the critic input $\mathbf{y}_n^t = [e_n^t, h_n^t, \iota_n]$. The centralized critic with parameters $\phi$ outputs the action-value vector of scheduling agent $n$ over all candidate actions:
\begin{equation}\label{eq:action_value_vector}
\mathbf{q}_\phi(\tilde{s}^t, n) = f_\phi(\mathbf{y}_n^t) \in \mathbb{R}^{|\mathcal{A}_n|},
\end{equation}
where $f_\phi(\cdot)$ denotes the per-agent value head of the critic. The scalar action-value function is defined as
\begin{equation}\label{eq:scalar_q}
Q_\phi(\tilde{s}^t, n, a) \triangleq [\mathbf{q}_\phi(\tilde{s}^t, n)]_a,
\end{equation}
where $[\cdot]_a$ denotes the vector entry corresponding to action $a$. To reduce training instability caused by Q-value overestimation, A-MADiff adopts a double critic structure, denoted by $Q_{\phi_1}$ and $Q_{\phi_2}$. When computing target values and updating the actors, A-MADiff uses the smaller of the two critic outputs:
\begin{equation}\label{eq:double_critic_min}
Q_{\phi}^{\min}(\tilde{s}^t, n, a) = \min_{j \in \{1,2\}} Q_{\phi_j}(\tilde{s}^t, n, a).
\end{equation}
The target counterpart $Q_{\bar{\phi}}^{\min}$ is defined
analogously over the target critics $Q_{\bar{\phi}_1}$ and
$Q_{\bar{\phi}_2}$.

\subsection{SAC Objective}\label{subsec:sac_objective}

Building on the per-agent marginal value estimation provided by the attention-guided critic, A-MADiff optimizes the diffusion-based actors and the centralized critics through entropy-regularized off-policy learning. We denote the target actor parameters as $\bar{\theta}$ and the target critic parameters as $\bar{\phi}_1$ and $\bar{\phi}_2$. Each experience sample from a mini-batch $\mathcal{B}$ drawn from the replay buffer $\mathcal{D}$ is defined as
\begin{equation}\label{eq:experience_tuple}
\zeta = (\tilde{s}^t, \mathbf{a}^t, \mathbf{r}^t, \tilde{s}^{t+1}, \mathbf{d}^t, \mathcal{F}^t, \mathcal{F}^{t+1}),
\end{equation}
where $\mathbf{a}^t = \{a_n^t\}_{n \in \mathcal{N}}$, $\mathbf{r}^t = \{r_n^t\}_{n \in \mathcal{N}}$, $\mathbf{d}^t = \{d_n^t\}_{n \in \mathcal{N}}$, $\mathcal{F}^t = \{\mathcal{F}_n^t\}_{n \in \mathcal{N}}$ collects the structurally feasible action sets of all agents, and $d_n^t$ is the terminal indicator variable for agent $n$ at this transition. The Temporal-Difference (TD) target of A-MADiff adopts per-agent reward decomposition and is conditioned on the joint local observation through the centralized critic, expressed as
\begin{equation}\label{eq:td_target}
y_n^t = r_n^t + \gamma (1 - d_n^t) V_{\bar{\phi}}(\tilde{s}^{t+1}, n),
\end{equation}
where the soft state value is given by
\begin{multline}\label{eq:soft_value}
V_{\bar{\phi}}(\tilde{s}^{t+1}, n) = \sum_{a \in \mathcal{F}_n^{t+1}} \bar{\pi}_n^{t+1}(a) 
\\ \times \left[ Q_{\bar{\phi}}^{\min}(\tilde{s}^{t+1}, n, a) - \alpha_{\mathrm{ent}} \log \bar{\pi}_n^{t+1}(a) \right].
\end{multline}
Here, $\bar{\pi}_n^{t+1}(a) \triangleq \pi_{\bar{\theta}}(a \mid o_n^{t+1}, \mathcal{F}_n^{t+1})$ denotes the masked target policy of scheduling agent $n$, and $\alpha_{\mathrm{ent}}$ denotes the entropy coefficient that controls exploration intensity. The critic is updated by minimizing the mean-squared TD error of the double critics, which is expressed as
\begin{equation}\label{eq:critic_loss}
\mathcal{L}_Q(\phi_1, \phi_2) = \frac{1}{|\mathcal{B}| N} \sum_{\zeta \in \mathcal{B}} \sum_{n=1}^{N} \sum_{j=1}^{2} \left( Q_{\phi_j}(\tilde{s}^t, n, a_n^t) - y_n^t \right)^2.
\end{equation}

The actor objective is to increase the selection probability of high-value actions over the structurally feasible action set while maintaining sufficient exploration. To avoid computing $\log 0$ for masked actions, the actor loss sums only over the current structurally feasible action set and is defined as
\begin{multline}\label{eq:actor_loss}
\mathcal{L}_\pi(\theta) = \frac{1}{|\mathcal{B}| N} \sum_{\zeta \in \mathcal{B}} \sum_{n=1}^{N} \sum_{a \in \mathcal{F}_n^t} \pi_n^t(a) \\
\times \left[ \alpha_{\mathrm{ent}} \log \pi_n^t(a) - Q_{\phi}^{\min}(\tilde{s}^t, n, a) \right],
\end{multline}
where $\pi_n^t(a) \triangleq \pi_\theta(a \mid o_n^t, \mathcal{F}_n^t)$ is the masked policy of scheduling agent $n$ defined in~(\ref{eq:masked_policy}). Minimizing~(\ref{eq:actor_loss}) shifts probability mass from low-value to high-value feasible actions.

To adaptively adjust the exploration intensity, A-MADiff further updates the entropy coefficient. The current policy entropy is computed as
\begin{equation}\label{eq:entropy}
\mathcal{H}_\pi = -\frac{1}{|\mathcal{B}| N} \sum_{\zeta \in \mathcal{B}} \sum_{n=1}^{N} \sum_{a \in \mathcal{F}_n^t} \pi_n^t(a) \log \pi_n^t(a).
\end{equation}
Given a target entropy $\bar{\mathcal{H}}$, the optimization objective of the entropy coefficient is given by
\begin{equation}\label{eq:alpha_loss}
\mathcal{L}_{\alpha_{\mathrm{ent}}} = \log \alpha_{\mathrm{ent}} \, (\mathcal{H}_\pi - \bar{\mathcal{H}}).
\end{equation}
Finally, A-MADiff updates the target networks via a soft update:
\begin{equation}\label{eq:soft_update}
\bar{\theta} \leftarrow \tau_{\mathrm{up}} \theta + (1 - \tau_{\mathrm{up}}) \bar{\theta}, \quad
\bar{\phi}_j \leftarrow \tau_{\mathrm{up}} \phi_j + (1 - \tau_{\mathrm{up}}) \bar{\phi}_j,
\end{equation}
where $j \in \{1, 2\}$, and $\tau_{\mathrm{up}} \in (0, 1]$ denotes the target network soft update coefficient.

\begin{algorithm}[t]
\caption{A-MADiff Training and Decentralized Execution}
\label{alg:amad_training}
\begin{algorithmic}[1]
\STATE \textbf{Initialize} joint diffusion-based decentralized policy $\pi_\theta$, double critics $Q_{\phi_1}$, $Q_{\phi_2}$, target networks $\pi_{\bar{\theta}}$, $Q_{\bar{\phi}_1}$, $Q_{\bar{\phi}_2}$, replay buffer $\mathcal{D}$, and the entropy coefficient $\alpha_{\mathrm{ent}}$
\STATE \textbf{// Centralized Training Phase}
\FOR{episode $= 1, 2, \ldots, E_{\max}$}
    \STATE Reset environment, obtain initial observations $\{o_n^0\}_{n \in \mathcal{N}}$
    \FOR{decision step $t = 0, 1, \ldots, T-1$}
        \FOR{each scheduling agent $n \in \mathcal{N}$}
            \STATE Construct $\mathcal{F}_n^t$ from $o_n^t$ via~(\ref{eq:feasibility})
            \STATE Sample $x_{n,T_d}^{t} \sim \mathcal{N}(\mathbf{0}, \mathbf{I})$
            \FOR{$\ell = T_d, T_d{-}1, \ldots, 1$}
                \STATE Compute $x_{n,\ell-1}^{t}$ via~(\ref{eq:noise_pred})--(\ref{eq:reverse_sample})
            \ENDFOR
            \STATE Obtain $z_n^t = x_{n,0}^{t}$, compute $\pi_\theta(a \mid o_n^t, \mathcal{F}_n^t)$ via~(\ref{eq:masked_policy})
            \STATE Sample action $a_n^t \sim \pi_\theta(\cdot \mid o_n^t, \mathcal{F}_n^t)$
        \ENDFOR
        \STATE Execute joint action $\mathbf{a}^t$, observe $\mathbf{r}^t$, $\tilde{s}^{t+1}$, $\mathbf{d}^t$, $\mathcal{F}^{t+1}$
        \STATE Store $(\tilde{s}^t, \mathbf{a}^t, \mathbf{r}^t, \tilde{s}^{t+1}, \mathbf{d}^t, \mathcal{F}^t, \mathcal{F}^{t+1})$ in $\mathcal{D}$
    \ENDFOR
    \STATE Sample mini-batch $\mathcal{B} \sim \mathcal{D}$
    \STATE Update the critics $Q_{\phi_1}, Q_{\phi_2}$ by minimizing~(\ref{eq:critic_loss})
    \STATE Update the policy $\pi_\theta$ by minimizing~(\ref{eq:actor_loss})
    \STATE Update the entropy coefficient $\alpha_{\mathrm{ent}}$ by minimizing~(\ref{eq:alpha_loss})
    \STATE Soft-update target networks via~(\ref{eq:soft_update})
\ENDFOR
\STATE \textbf{// Decentralized Execution Phase}
\FOR{each decision step $t$}
    \FOR{each scheduling agent $n \in \mathcal{N}$}
        \STATE Observe $o_n^t$ and construct $\mathcal{F}_n^t$
        \STATE Select $a_n^t$ using trained $\pi_\theta$ and $\mathcal{F}_n^t$ via~(\ref{eq:masked_policy})
    \ENDFOR
\ENDFOR
\end{algorithmic}
\end{algorithm}

\subsection{Training and Decentralized Execution of A-MADiff}\label{subsec:training_exec}

Algorithm~\ref{alg:amad_training} summarizes the training and execution procedure of A-MADiff, following the CTDE paradigm. The training phase first initializes the diffusion-based actors, the double centralized critics, the corresponding target networks, and the replay buffer. At each decision step during training, every scheduling agent constructs its structurally feasible action set $\mathcal{F}_n^t$ from the local observation $o_n^t$, and its diffusion-based actor generates an orchestration action through masked reverse denoising. The environment then updates task states, queue states, ASP resource occupancy, and execution outcomes based on the joint action, and stores the transition in the replay buffer. At each training step, a mini-batch is sampled from the replay buffer, and the actor, critics, and entropy coefficient are updated in an off-policy manner. This off-policy structure improves sample efficiency and helps mitigate training fluctuations caused by dynamic task arrivals and multi-agent interactions~\cite{zhao2022multi, du2024diffusion}. During training, the centralized critic uses the joint local observation to evaluate the long-term value of each scheduling agent's local action and provides centralized value guidance for the diffusion-based actors through TD error updates. After training is complete, the critic is no longer involved in online decision-making. Each scheduling agent independently selects a local execution, peer offloading, or queue deferral action based solely on its own local observation $o_n^t$ and structurally feasible action set $\mathcal{F}_n^t$. All scheduling agents share one set of actor parameters, while their decisions differ through heterogeneous local observations and structurally feasible action sets.

The online execution overhead of A-MADiff primarily comes from the multi-step denoising of the diffusion-based actors and the masked softmax. We denote the action dimension as $A = |\mathcal{A}_n|$, the actor hidden dimension as $d_h$, and the number of reverse denoising steps as $T_d$. Since the noise prediction network adopts a Multi-Layer Perceptron (MLP) structure with residual connections, the actor inference complexity per scheduling agent is $\mathcal{O}(T_d d_h^2 + T_d A d_h)$, and the masked softmax complexity is $\mathcal{O}(A)$. Therefore, the total decentralized execution complexity across all $N$ scheduling agents is $\mathcal{O}( N T_d d_h^2 + N T_d A d_h )$. This complexity involves only the decentralized actor inference and does not require the centralized critic. The additional training overhead primarily comes from the attention-guided centralized critic. For a mini-batch $\mathcal{B}$, with single-agent observation dimension $d_o$, the main complexity of the observation encoding and the attention-based critic can be approximated as $\mathcal{O}\!\left( |\mathcal{B}| N d_o d_h + |\mathcal{B}| N^2 d_h + |\mathcal{B}| N d_h^2 \right)$. The three terms correspond to the observation encoding across $N$ agents, the pairwise attention computation among all agents, and the per-agent value head forward pass, respectively. Since the centralized critic is used only during training, this complexity does not increase the online execution overhead.


\section{Experimental Results}\label{sec:experiments}


\subsection{Experimental Setup}\label{subsec:exp_setup}

\begin{table}[!t]
\centering
\caption{Simulation parameters.}
\label{tab:sim_params}
\footnotesize
\setlength{\tabcolsep}{3.5pt}
\begin{tabular}{llr}
\toprule
\rowcolor{gray!6}
\rule[-1.0ex]{0pt}{3.4ex}\textbf{Category} & \textbf{Parameters} & \textbf{Values} \\
\midrule
\multirow{4}{*}{System}
 & Scheduling agents $N$ & $20$ \\
 & ASPs per agent $|\mathcal{K}_n|$ & $5$ \\
 & Queue capacity $Q_{\max}$ & $50$ \\
 & Max offloading hops $H_{\max}$ & $2$ \\
\midrule
\multirow{5}{*}{Task}
 & Arrival rate $\lambda_n$ (per step) & $0.3$ \\
 & Pre-generated task pool per agent & $500$ \\
 & Workload $w_i$ & $\{50,100,150,300\}$ \\
 & Memory requirement $m_i$ (GB) & $\{8,16,24\}$ \\
 & Data payload $D_i$ (MB) & $\mathcal{U}(1,50)$ \\
\midrule
\multirow{5}{*}{ASP}
 & Inference rate $p_{n,k}^{\mathrm{base}}$ (steps/s) & $\{10,50,100,200\}$ \\
 & Memory capacity $M_{n,k}^{\mathrm{cap}}$ (GB) & $\{12,24,40,80\}$ \\
 & Rate perturbation $\varepsilon$ & $\mathcal{U}(-0.2,0.2)$ \\
 & Bandwidth $B_{n,n'}$ (MB/s) & $20$ \\
 & Link latency $\Delta_{n,n'}$ (s) & $\mathcal{U}(0.1,1.0)$ \\
\midrule
\multirow{3}{*}{Reward}
 & Latency weight $\eta_T$ & $0.1$ \\
 & Deferral cost $R_{\mathrm{def}}$ & $-0.5$ \\
 & Failure penalty $R_{\mathrm{fail}}$ & $-5.0$ \\
\midrule
\multirow{7}{*}{Training}
 & Episodes / Steps per episode & $2000$ / $200$ \\
 & Denoising steps $T_d$ & $20$ \\
 & Hidden dim / Attention heads & $256$ / $4$ \\
 & Actor / Critic learning rate & $10^{-4}$ / $3{\times}10^{-4}$ \\
 & Replay buffer / Batch size & $2{\times}10^5$ / $1024$ \\
 & Discount factor $\gamma$ & $0.99$ \\
 & Soft update coefficient $\tau_{\mathrm{up}}$ & $0.005$ \\
\bottomrule
\end{tabular}
\end{table}

We evaluate A-MADiff in a simulated memory-constrained mobile AIGC network whose parameters are listed in Table~\ref{tab:sim_params}. The system consists of $N = 20$ scheduling agents, each managing $5$ heterogeneous ASPs. Task arrivals follow a Poisson process with exponentially distributed inter-arrival times~\cite{du2024diffusion, liu2025qos}, and only the tasks arriving within the episode horizon enter the system. The runtime inference rate of each ASP is $\mu_{n,k}^t = p_{n,k}^{\mathrm{base}}(1 + \varepsilon)$, where $p_{n,k}^{\mathrm{base}}$ denotes the nominal inference rate of ASP $k$ and $\varepsilon$ is a uniform random perturbation~\cite{li2024llm, qu2025mobile}. All methods share the same structurally feasible action set, runtime admission check, and reward function. The service utility follows the human-aware modeling approach in~\cite{du2024diffusion}, normalized to $u_{n,k}^i \in [0,1]$.

A-MADiff is compared against three heuristic baselines, i.e., Random, RoundRobin, and Greedy, as well as four MADRL-based baselines, i.e., Multi-Agent Deep Deterministic Policy Gradient (MADDPG)~\cite{lowe2017multi}, Multi-Agent Proximal Policy Optimization (MAPPO)~\cite{yu2022surprising}, Multi-Agent SAC (MASAC)~\cite{haarnoja2018soft}, and MADiff\footnote{MADiff is an ablation variant of A-MADiff that removes the attention-guided critic. It is distinct from the offline framework by Zhu et al.~\cite{zhu2024madiff}, which uses supervised denoising with a U-Net architecture.}. Each method is evaluated over $20$ independent episodes. During training, the policies are additionally evaluated at regular intervals on separate evaluation episodes, and the resulting test reward reflects the cumulative reward achieved by the learned policy at that point.

\begin{table*}[t]
\caption{Performance comparisons of A-MADiff and baselines. Bold values indicate the best result in each column.}
\label{tab:performance}
\centering
\renewcommand{\arraystretch}{1.3}
\setlength{\tabcolsep}{3.5pt}
\begin{tabular}{c l c c c c c}
\toprule
\rowcolor{gray!6}
 & \textbf{Methods}
 & \textbf{Cumulative Rewards}
 & \textbf{Effective Success Rates}
 & \textbf{Rewards per Completed Task}
 & \textbf{Negative Completion Rates}
 & \textbf{OOM Rates} \\
\midrule
\multirow{3}{*}{\shortstack{Heuristic\\Methods}}
 & Random     & $-407.78 \pm 19.96$ & $8.18\%$  & $-0.0577$ & $42.46\%$ & $5.78\%$ \\
 & RoundRobin & $-373.35 \pm 19.41$ & $7.94\%$  & $-0.0923$ & $45.97\%$ & $4.85\%$ \\
 & Greedy     & $281.84 \pm 19.13$  & $77.71\%$ & $0.2414$  & $22.29\%$ & $2.07\%$ \\
\midrule
\multirow{5}{*}{\shortstack{MADRL\\Methods}}
 & MADDPG & $355.81 \pm 19.32$ & $87.31\%$ & $0.3260$ & $12.68\%$ & $1.56\%$ \\
 & MAPPO  & $422.38 \pm 31.35$ & $90.87\%$ & $0.4036$ & $8.94\%$  & $1.04\%$ \\
 & MASAC  & $476.58 \pm 16.59$ & $91.29\%$ & $0.4278$ & $8.68\%$  & $1.35\%$ \\
 & MADiff & $508.77 \pm 18.06$ & $95.27\%$ & $0.4554$ & $4.73\%$  & $1.23\%$ \\
 & \cellcolor{oursgreen}\textbf{A-MADiff}
 & \cellcolor{oursgreen}$\mathbf{658.25 \pm 25.20}$
 & \cellcolor{oursgreen}$\mathbf{98.15\%}$
 & \cellcolor{oursgreen}$\mathbf{0.6232}$
 & \cellcolor{oursgreen}$\mathbf{1.77\%}$
 & \cellcolor{oursgreen}$\mathbf{1.01\%}$ \\
\bottomrule
\end{tabular}
\end{table*}

\begin{figure*}[!t]
\vspace{-16pt}
\centering
\subfloat[Test rewards for A-MADiff and representative
\mbox{heuristic methods}.]{%
\begin{minipage}[t]{0.32\textwidth}
  \centering
  \includegraphics[width=\textwidth,trim=0 10bp 0 8bp,clip]
  {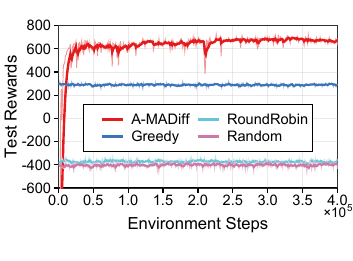}%
\end{minipage}%
\label{fig:baseline_reward}}%
\hfill
\subfloat[Test rewards for A-MADiff and representative
\mbox{MADRL methods}.]{%
\begin{minipage}[t]{0.32\textwidth}
  \centering
  \includegraphics[width=\textwidth,trim=0 10bp 0 8bp,clip]
  {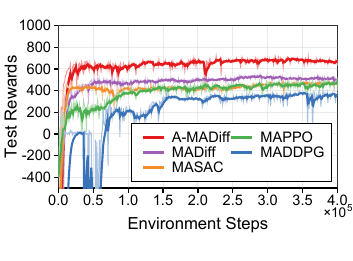}%
\end{minipage}%
\label{fig:convergence}}%
\hfill
\subfloat[Environment steps to match the best-performing
heuristic method.]{%
\begin{minipage}[t]{0.32\textwidth}
  \centering
  \includegraphics[width=\textwidth,trim=0 10bp 0 8bp,clip]
  {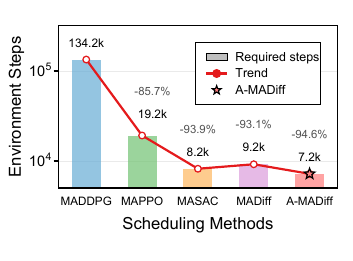}%
\end{minipage}%
\label{fig:greedy_steps}}
\caption{Training performance of all evaluated methods.
The heuristic baselines provide constant reference levels, the MADRL methods are evaluated at regular intervals on separate evaluation episodes during training, and the vertical axis in~(c) uses a logarithmic~scale.}
\label{fig:training_behavior}
\end{figure*}

In addition to cumulative reward, we report four metrics. We denote the number of arrived tasks as $N_{\mathrm{arr}}$, the number of completed tasks as $N_{\mathrm{comp}}$, and the completed task set as $\mathcal{C}_{\mathrm{comp}}$. The cumulative path reward of a completed task $i \in \mathcal{C}_{\mathrm{comp}}$ is denoted as $r_i$, which sums the immediate rewards incurred at each decision step along the scheduling path of task $i$, from arrival to execution completion. The number of local execution failures caused by insufficient GPU memory at runtime is denoted by $N_{\mathrm{oom}}$. A task counts as completed only when it is executed by an ASP, not when it enters a queue. Unlike a completion rate $N_{\mathrm{comp}}/N_{\mathrm{arr}}$, which does not differentiate completions with positive versus non-positive cumulative reward, the effective success rate $\sum_{i \in \mathcal{C}_{\mathrm{comp}}} \mathbb{I}(r_i > 0) / N_{\mathrm{arr}}$ measures the fraction of arrived tasks that are completed with positive cumulative reward. The reward per completed task $\sum_{i \in \mathcal{C}_{\mathrm{comp}}} r_i / N_{\mathrm{comp}}$ measures average completion quality, while the negative completion rate $\sum_{i \in \mathcal{C}_{\mathrm{comp}}} \mathbb{I}(r_i \le 0) / N_{\mathrm{comp}}$ quantifies the fraction of completed tasks with non-positive reward. The OOM rate $N_{\mathrm{oom}}/N_{\mathrm{arr}}$ measures how often such failures occur among the arrived tasks.

\begin{figure}[!t]
\centering
\includegraphics[width=0.84\columnwidth,trim=0 8bp 0 4bp,clip]{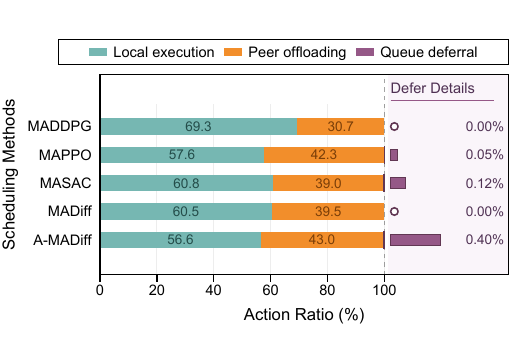}
\vspace{-6pt}
\caption{Action composition of different MADRL methods during evaluation, showing the ratios of local execution, peer offloading, and queue deferral.}
\label{fig:action_composition}
\end{figure}

\begin{figure*}[t]
\centering
\includegraphics[width=\textwidth]{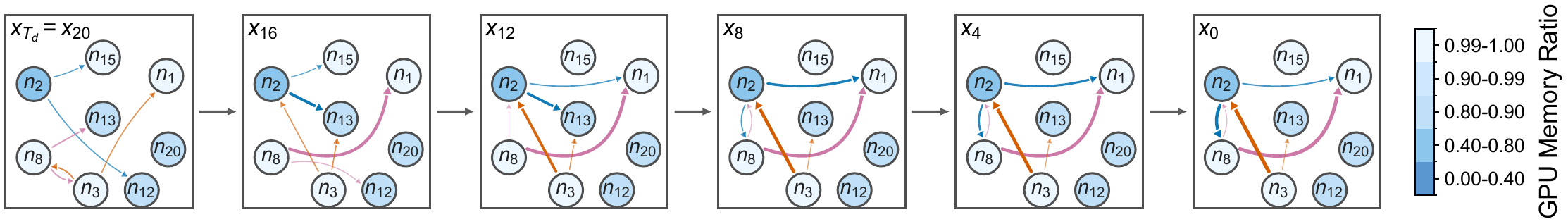}
\caption{Reverse denoising process of the diffusion-based actors within one joint scheduling decision at decision step $t = 58$. Arrow width and opacity encode the probabilities obtained by applying the masked softmax in (\ref{eq:masked_policy}) to the intermediate logit vector $x_{n,\ell}^{t}$ at denoising step $\ell \in \{20, 16, 12, 8, 4, 0\}$, and only the two largest offloading candidates of each agent are drawn. Node color encodes the available GPU memory ratio in (\ref{eq:mem_ratio}). Blue, purple, and orange arrows denote the preferences of scheduling agents $n_2$, $n_8$, and $n_3$, respectively, and solid arrows in the final panel mark the selected actions.}
\label{fig:denoising_process}
\end{figure*}

\begin{figure}[t]
\centering
\setlength{\abovecaptionskip}{3pt}
\includegraphics[width=0.85\columnwidth,trim=0 5bp 0 6bp,clip]{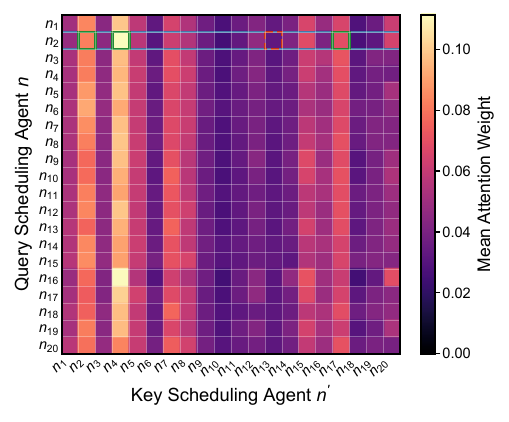}
\caption{Mean attention weights of the centralized critic at decision step $t = 7$, averaged over the four attention heads and the two critics. Rows and columns correspond to the query agent $n$ and the key agent $n'$ in (\ref{eq:multi_head_attention}). Green boxes mark the three largest weights in the row of $n_2$, and the orange dashed box marks the offloading target selected by the actor.}
\label{fig:attention_heatmap}
\end{figure}

\subsection{Overall Performance and Convergence}\label{subsec:overall_performance}

Figure~\ref{fig:training_behavior} and Table~\ref{tab:performance} present the overall performance comparison. Among the heuristic baselines in Fig.~\ref{fig:baseline_reward}, Random and RoundRobin yield negative cumulative rewards with effective success rates below $9\%$, confirming that uncoordinated scheduling fails under GPU memory constraints. Greedy substantially outperforms both by selecting the highest immediate-reward action at each step. However, its elevated OOM rate and negative completion rate indicate that single-step optimization ignores cross-step GPU memory occupancy and queue dynamics. All four MADRL baselines surpass Greedy, confirming that learned policies capture long-horizon orchestration value. Among them, MASAC outperforms MADDPG and MAPPO through entropy-regularized off-policy learning. Replacing the MLP actor in MASAC with a diffusion-based actor further improves the cumulative reward and reduces the negative completion rate by nearly half. This result suggests that the diffusion-based actors capture state-dependent multi-modal action preferences more effectively than a single-pass MLP actor. Moreover, A-MADiff improves the cumulative reward over MADiff by $29.4\%$ and over MASAC by $38.1\%$. 

As shown in Table~\ref{tab:performance}, A-MADiff achieves the highest effective success rate and reward per completed task among all methods, while attaining the lowest negative completion rate and the lowest OOM rate. Since the structural mask already excludes physically infeasible local execution, OOM events arise only from temporary runtime memory shortfalls, and learned policies further reduce these failures through the penalty signal. The gap between MADiff and A-MADiff isolates the contribution of the attention-guided centralized critic. By selectively aggregating cross-agent states during training, the critic helps the actor distinguish which neighboring memory and load conditions are most relevant to the current offloading decision. Figure~\ref{fig:convergence} shows the convergence trends of the learning-based methods. A-MADiff converges to a higher reward plateau than all baselines and maintains stable training throughout the $4 \times 10^5$ environment steps. The clear method ordering across the training trajectory is consistent with the progressive addition of entropy regularization in MASAC, diffusion-based action modeling in MADiff, and attention-guided value estimation in A-MADiff.

\subsection{Learning Efficiency and Action Composition}\label{subsec:learning_action}
 
\subsubsection{Learning efficiency}
 
Figure~\ref{fig:greedy_steps} shows the number of environment steps required by each MADRL-based method to reach the test reward level of the Greedy baseline. A-MADiff reaches this level with the fewest steps among all methods. Notably, MADiff requires slightly more steps than MASAC despite achieving a higher final reward, which is consistent with the expectation that the multi-step reverse denoising process produces more dispersed early exploration. By adding the attention-guided centralized critic, A-MADiff reduces the required steps by $21.7\%$ compared with MADiff, suggesting that selective cross-agent value estimation partially offsets the dispersed early exploration of the diffusion-based actors.
 
\subsubsection{Action composition}
 
Figure~\ref{fig:action_composition} shows the action composition of different MADRL methods during evaluation. Across all these methods, the offloading ratio broadly increases with the cumulative reward, with MADDPG and A-MADiff marking the lowest and highest ends of both orderings. MADDPG, which uses a deterministic actor, concentrates the largest fraction of actions on local execution. Stochastic-policy baselines, i.e., MAPPO, MASAC, and MADiff, distribute actions more evenly between local execution and offloading. MASAC and MADiff exhibit similar macroscopic action compositions, yet MADiff achieves a higher reward per completed task and a lower negative completion rate. This result indicates that the diffusion-based actors enable finer-grained and state-dependent action selection within comparable macroscopic ratios. Alongside the highest offloading ratio, A-MADiff attains the lowest local execution ratio and retains a small but non-zero deferral ratio. The high offloading ratio indicates that the policy actively distributes tasks across neighboring ASPs to reduce local memory pressure. The non-zero deferral ratio suggests that the policy introduces limited scheduling buffering, avoiding forced local execution or low-value offloading.

\subsection{Interpretability Analysis}\label{subsec:interpretation}

\subsubsection{Reverse denoising of the diffusion-based actor}

As illustrated in Fig.~\ref{fig:denoising_process}, we trace one joint decision at decision step $t = 58$, where the pending task at scheduling agent $n_2$ requires $24$ GB of GPU memory, the largest available GPU memory at runtime among the local ASPs of $n_2$ is $12$ GB, and the normalized queue length reaches $0.98$. Selecting local execution would therefore fail the runtime admission check in (\ref{eq:runtime_admissible}) and incur the failure penalty, although the structural mask in (\ref{eq:feasibility}) retains it as a physically feasible option. As the denoising step decreases from $T_d = 20$ to $0$, the dispersed preferences of $n_2$ concentrate on $n_8$ with a final probability of $0.8020$, whereas $n_{15}$ receives only $3.78 \times 10^{-7}$ despite an available memory ratio close to $1$ and a $40$ GB ASP, since $n_8$ offers higher inference capability and lower communication latency. The trained diffusion-based actor therefore weighs memory feasibility, inference capability, queue load, communication conditions, and long-term value jointly, and avoids the runtime-infeasible local mode without perfect knowledge of per-ASP memory availability.

\begin{figure*}[!t]
\centering
\includegraphics[width=0.585\textwidth,trim=0 1.5bp 0 1.5bp,clip]%
{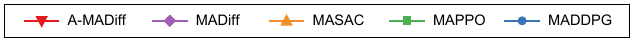}\\[-1pt]
\subfloat[Impact of the GPU memory capacity scale.]{%
\includegraphics[width=0.32\textwidth,trim=0 0 0 6bp,clip]%
{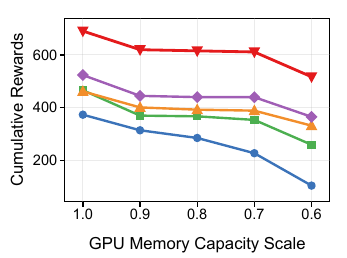}%
\label{fig:robustness_capacity}}
\hfil
\subfloat[Impact of the GPU memory requirement scale.]{%
\includegraphics[width=0.32\textwidth,trim=0 0 0 6bp,clip]%
{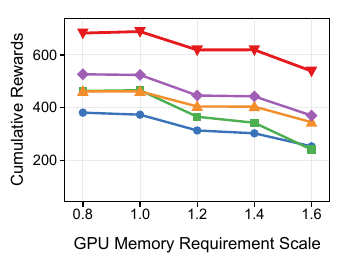}%
\label{fig:robustness_requirement}}%
\hfil
\subfloat[Impact of the task arrival scale.]{%
\includegraphics[width=0.32\textwidth,trim=0 0 0 6bp,clip]%
{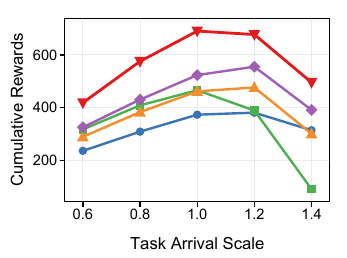}%
\label{fig:robustness_arrival}}
\caption{Robustness comparison of the cumulative reward under three scaling dimensions. The GPU memory capacity scale uniformly scales the GPU memory capacity of all ASPs, the GPU memory requirement scale uniformly scales the GPU memory requirements of all tasks, and the task arrival scale uniformly scales the task arrival rate at all scheduling agents. Each experiment varies one dimension while the others remain at their default values.}
\label{fig:robustness}
\end{figure*}

\subsubsection{Attention weights of the centralized critic}

As shown in Fig.~\ref{fig:attention_heatmap}, at an independent decision step $t = 7$, scheduling agent $n_2$ schedules a task requiring $24$ GB of GPU memory while the largest available GPU memory among its local ASPs is $16$ GB. The three largest weights in the row of $n_2$ fall on $n_4$ ($0.1113$), $n_2$ itself ($0.0813$), and $n_{17}$ ($0.0684$), all clearly above the uniform level of $1/N = 0.05$, whereas the actor routes the task to $n_{13}$. The attended agents identify the cross-agent context most relevant to long-term value estimation rather than mirroring the routing target. The critic thus replaces uniform aggregation with selective, decision-relevant weighting, consistent with the gain of A-MADiff over MADiff in Table~\ref{tab:performance}.

\subsection{Robustness Analysis}\label{subsec:robustness}

\subsubsection{GPU memory pressure}

Figure~\ref{fig:robustness_capacity} shows the cumulative reward under different GPU memory capacity scales, where a smaller scale uniformly shrinks the GPU memory capacity of all ASPs while task memory requirements remain unchanged. As the capacity scale decreases from $1.0$ to $0.6$, the cumulative rewards of all methods decline, confirming that reduced available GPU memory increases the risk of admission failures and local resource contention. Figure~\ref{fig:robustness_requirement} evaluates memory pressure from the demand side, where a larger scale uniformly inflates the GPU memory requirements of all tasks while ASP capacities remain unchanged, tightening the admission condition for every task. As the requirement scale increases from $1.0$ to $1.6$, the cumulative rewards decline again, and A-MADiff degrades most gracefully under both forms of memory pressure. These results support the modeling choice that treats GPU memory as a hard feasibility constraint: The policy needs to actively adjust offloading behavior as the memory margin shrinks, rather than relying solely on computation capacity.

\subsubsection{Workload intensity variation}

Figure~\ref{fig:robustness_arrival} shows the cumulative reward under different task arrival rate scales. As the arrival rate increases, the number of concurrent tasks, waiting queue pressure, and ASP memory contention increase accordingly, causing the cumulative rewards of all methods to first increase and then decline. Under low to moderate workload conditions, A-MADiff achieves higher rewards through more effective peer offloading and task matching. As the arrival rate continues to increase, all methods experience reward declines due to task congestion, yet A-MADiff maintains the highest cumulative reward across the tested range. This result indicates that A-MADiff exhibits comparatively stable orchestration performance under varying workload intensities.

Overall, A-MADiff attains the best value on every metric in Table~\ref{tab:performance}, while retaining the highest cumulative reward across all three scaling dimensions. 


\section{Conclusion}\label{sec:conclusion}

We have studied GPU memory-aware task orchestration in mobile AIGC networks, where insufficient GPU memory causes outright task execution failure rather than latency degradation. First, we have formulated the AIGC task orchestration process as a cooperative Dec-POMDP that explicitly treats the compatibility between the GPU memory requirement of each AIGC task and the GPU memory available at the serving ASP as a hard feasibility constraint. To solve the formulated problem, we have proposed A-MADiff under the CTDE paradigm. A-MADiff combines diffusion-based decentralized actors for discrete action preference generation with an attention-guided centralized critic for selective cross-agent value estimation. Numerical results demonstrate that A-MADiff improves the cumulative reward by $29.4\%$ over the state-of-the-art baseline, while achieving the highest effective success rate and the lowest OOM rate among all evaluated methods. For future work, we will extend the proposed framework to the orchestration of AI agent tasks, and validate it on physical edge testbeds with heterogeneous GPU configurations.


\bibliographystyle{IEEEtran}
\bibliography{references}


\end{document}